\documentclass[12pt]{iopart}

\usepackage{iopams}
\usepackage{graphicx}

\newcommand{\pspoint}{{x}}

\newcommand {\bopQ}{Q}
\newcommand {\bopP}{P}

\newcommand {\bindex} {\alpha}

\newcommand {\bopps}{\mathcal S}

\newcommand {\cas}{{\mathcal L}}

\newcommand {\dif}{{\rm d}}

\newcommand {\F}{F}

\newcommand {\HTOT} { H _{{\rm tot}}}

\newcommand {\HS} { H _{{\rm s}}}

\newcommand{\opA}{A}

\newcommand {\opB} {B}

\newcommand{\hA}{A}

\newcommand {\iu}{{\rm i}}

\newcommand {\ord} {\sigma}
\newcommand {\opS}{S}

\newcommand {\psF} {W}

\newcommand {\sws}{\Delta}

\newcommand {\T}{T}

\newcommand {\tL} {{\widehat {\Lambda}}}

\newcommand {\vl}{|}

\newcommand {\W} {W_{\varrho}}

\newcommand {\coef}{\eta}
\newcommand {\tcoef}{\tilde{\eta}}

\begin{document}

\title[Bopp operators and phase-space spin dynamics]{Bopp operators
and phase-space spin dynamics: Application to  rotational quantum
brownian motion}

\author{
D. Zueco$^1$ and I. Calvo$^2$\\[10pt] 
}

\address{
$^1$
Departamento de F\'{\i}sica de la Materia Condensada e\\
Instituto de Ciencia de Materiales de Arag\'on\\
C.S.I.C.--Universidad de Zaragoza\\
E-50009 Zaragoza, Spain
\\
E-mail: {\tt zueco@unizar.es}
\\[10pt]
$^2$
Laboratorio Nacional de Fusi\'on\\
Asociaci\'on EURATOM-CIEMAT\\
E-28040 Madrid, Spain\\
E-mail: {\tt ivan.calvo@ciemat.es}
}

\date{\today}

\begin{abstract}
For non-relativistic spinless particles, Bopp operators give an elegant
and simple way to compute the dynamics of quasiprobability
distributions in the phase space formulation of Quantum Mechanics. In
this work, we present a generalization of Bopp operators for spins and
apply our results to the case of open spin systems. This approach
allows to take the classical limit in a transparent way, recovering
the corresponding Fokker-Planck equation.
\end{abstract}



\section{Introduction}

The seed of the phase-space formulation of Quantum Mechanics dates
back to 1932, when Wigner introduced his famous quasi-probability
function in phase space \cite{wig32}:
\begin{eqnarray}
\nonumber
W_\psi(q,p)=\frac{1}{2\pi\hbar}\int_{-\infty}^\infty \dif u 
\;
\psi^*(q-  \case{1}{2} u)
\; \psi(q+\case{1}{2}u) e^{-\iu u p/\hbar}.
\end{eqnarray}
The relation between Wigner's function and Weyl's correspondence was
understood by Groenewold \cite{gro46} and fully developed by Moyal
\cite{moy49}, who established an independent formulation of Quantum
Mechanics in phase space for the canonical Poisson bracket
$\{q,p\}=1$. In this equivalent, `statistical' or hydrodynamical
formulation of Quantum Mechanics, the expectation value of an operator
is computed as the average of the corresponding function on the phase
space with the `probability' density given by the Wigner function.

In light of these results, it was suggestive to look for an autonomous
formulation of Quantum Mechanics in phase space involving only
classical functions and valid for any Poisson bracket. This is the
so-called {\it deformation quantization} program proposed by M. Flato
and collaborators in the 1970s \cite{BayFlaFroLicSte}. The problem
is stated as follows: given a smooth manifold $M$ with Poisson bracket
$\{\cdot,\cdot\}:C^\infty(M)\times C^\infty(M)\rightarrow
C^\infty(M)$, find an ${\mathbb R}[[\hbar]]$-bilinear associative
deformation of the point-wise product on $C^\infty(M)$, $\star$, such
that
$$f\star g -g\star f=\hbar\{f,g\}+O(\hbar^2),\ \forall f,g\in
C^\infty(M).$$

The existence of $\star$-products for any Poisson manifold was a
longstanding problem in Mathematics solved in 1997 by M. Kontsevich
\cite{Kon}. As a consequence of his more general Formality Theorem, he
showed that any Poisson manifold can be quantized by deformation and
classified the $\star$-products (see globalization aspects in
\cite{CatFelTom}).

One of the main advantages of the deformation quantization prescription (or
equivalently, the phase-space formulation of Quantum Mechanics)
resides in the quantum-classical transition problem. The classical
limit is obtained in a clear and mathematically rigorous way by taking
$\hbar\rightarrow 0$.

The phase-space formalism has been applied successfully to the
description of a spinless particle (\cite{hiletal84, lee95}), whose
phase space is $\mathbb R^2$ with the canonical Poisson bracket. More
complicated is the phase-space description of a spin, the phase space
being the sphere $S^2$. The phase space of a spin was given in terms
of the atomic (or spin) coherent states, \cite{areetal72, takshi75,
takshi76, aga81}, which in a sense generalizes the Cahill-Glauber
construction for the spinless case \cite{cahgla69}.

Although the results of Kontsevich guarantee the existence of a
$\star$-product for any classical system (i.e.,  for any Poisson
manifold), his explicit formula is rather complicated and, for
example, it is not obvious how to take advantage of the symmetries of
the system. V\'arilly and Gracia-Bond\'{\i}a (see \cite{vargra89} for
$SU(2)$ and \cite{FigGraVar} for any compact group) showed that the
appropriate setup for systems with symmetries is given by the {\it
Stratonovich-Weyl (SW) correspondence} \cite{str56} (see Section
\ref{Background}). In addition, they proved that this approach is
equivalent to the spin coherent state representation. These works were
generalized by Brif and Mann \cite{briman98, briman99}, establishing
the SW correspondence for systems with an arbitrary finite-dimensional
Lie-group symmetry.

In the present work we tackle the problem of the dynamics of spin
systems in the phase-space formulation and its classical
limit. Recently Klimov and Espinoza \cite{kliesp02} derived a
differential form for the star product in the spin case. On the other
hand, in the spinless case the evaluation of the star product becomes
simpler with the use of {\it Bopp operators} \cite{bop56}. We use
Klimov and Espinoza's results and work out the Bopp operators for
spins.  Takahashi and Shibata already found this generalization for
the specific cases of normal and anti-normal ordering
\cite{takshi75,takshi76}. Herein, we re-obtain their results and
generalize them for an arbitrary ordering.  In particular, we obtain
the Bopp operators for the important case of symmetric ordering.  We
will apply our results to show how the use of Bopp operators
simplifies the derivation of the dynamical equations for spins.

Finally, we deal with the problem of a system in contact with a
bosonizable bath as a model of quantum dissipation \cite{weiss}.
Classically, the effective dynamics of the system is described by
Langevin or Fokker-Planck equations \cite{zwanzig}. In the quantum
domain and under certain conditions (essentially weak coupling between
system and environment), the dynamics of the system can be formulated
in terms of a quantum master equation for the (reduced) density matrix
\cite{brepet}. For spinless particles the phase space transform of the
master equations yields the quantum generalization of the
Klein-Kramers equations \cite{calleg83pa}. In the sense of Caldeira
and Leggett, who pose the open system dynamics as a quantization
problem \cite{calleg83}, the phase-space formulation ``closes the
circle", giving a quantum version of the Fokker-Planck equations. With
the help of the Bopp operators for spins we easily obtain quantum
Fokker-Planck equations, recovering the corresponding classical
Fokker-Planck equations for the rotational brownian motion
\cite{gar2000} in the limit $\hbar\rightarrow 0$. This provides a
natural framework to link the classical and quantum theories of
dissipation.

\vskip 0.2cm

The paper is organized as follows:

\vskip 0.2cm

Section \ref{Background} is a brief survey on the phase-space
formulation of Quantum Mechanics from the point of view of the
Stratonovich-Weyl postulates. The general results are illustrated by
the example of the non-relativistic spinless particle, introducing the
definition of Bopp operators.

Section \ref{sec:spincase} deals with the phase-space formalism of
spin systems. We generalize the Bopp operators and tackle the problem of
the quantum-classical transition in these systems. In addition, we
work out in detail the particular case of quadratic Hamiltonians.

In Section \ref{sec:quantumME} we apply our results to open quantum
spin systems. We transform into phase space the density matrix
equation for a spin in contact with a thermal bath, obtaining quantum
Fokker-Planck equations. This transformation becomes simple with
the help of Bopp operators. We write explicitly the dissipative
equations for the linear Hamiltonian ({\it isotropic} spin) and compare it
with its spinless analogue, the damped harmonic oscillator.  Finally,
we take the classical limit for a general quantum Fokker-Planck
equation, recovering the corresponding classical equation.

\section{Phase-space Quantum Mechanics}\label{Background}

In this section we review some basic facts and results on phase-space
quantization. The material is now standard and we closely follow the
conventions and notation of \cite{briman98, briman99}.

\subsection{The SW postulates}
Consider a physical system possessing a group of symmetries described
by a finite-dimensional, connected and simply connected Lie group
$G$. We denote the phase space of the system by $\cal P$ and assume
that $G$ acts transitively on $\cal P$. That is, for any
$\pspoint_1,\pspoint_2\in\cal P$, there exists $g\in G$ such that
$g\cdot \pspoint_1=\pspoint_2$. Let $\cal H$ be the Hilbert space of
our system, ${\cal O}(\cal H)$ the set of operators on $\cal H$ and
${\cal U}:G\rightarrow {\cal O}(\cal H)$ an irreducible unitary
representation of $G$. The Stratonovich-Weyl correspondence
(\cite{str56}) is a $\sigma$-parameterized map $W^{(\sigma)}:{\cal
O}(\cal H)\rightarrow C^\infty(\cal P)$ satisfying the following
properties for any $A,\opB \in {\cal O}({\cal H})$:
\begin{enumerate}
\item Linearity:

\vskip0.2cm
$\hA \mapsto \psF_{A}^{(\ord)}$ is linear and bijective.
\vskip0.2cm

\item Reality:

\vskip0.2cm
$\psF_{A^\dagger}^{(\ord)}(\pspoint) = \psF_{A}^{(\ord)}(\pspoint)^{*}$, where
$\hA^{\dagger}$ is the adjoint of $\hA$.
\vskip0.2cm

\item Standardization: 
$$\int _{\cal P} \psF_{A}^{(\ord)}(\pspoint)\dif\mu(\pspoint) = \Tr A.$$
\item Traciality:
$$\int _{\cal P}\psF_{\opA}^{(\ord)}(\pspoint)
\psF_{\opB}^{(-\ord)}(\pspoint)\dif\mu(\pspoint) = \Tr(\opA\opB).$$
\item Covariance:
$$\psF_{A^g}^{(\ord)}(\pspoint) = \psF_{A}^{(\ord)}(g\cdot\pspoint), \ \forall
  g\in G.$$
\end{enumerate}
\noindent where $\mu$ is an invariant measure and $A^g:={\cal
U}(g^{-1})A{\cal U}(g)$.

As proved by Wigner {\it et al.} \cite{hiletal84} for spinless
particles, by V\'arilly and Gracia-Bond\'{\i}a for spins
\cite{vargra89} and recently by Brif and Mann \cite{briman98,
briman99} for systems with an arbitrary Lie-group symmetry, properties
(i)-(v) uniquely determine the SW correspondence.

Reality condition (ii) ensures that $W_A^{(\sigma)}$ is real whenever
$\hA$ is hermitian and property (iii) is simply a
normalization. Linearity and Traciality conditions are essential to
the formal interpretation of Quantum Mechanics as a statistical
theory. Let $\W^{(\ord)}$ be the image by the SW correspondence of the
density matrix of the system, $\varrho$. For any operator $A$,
$\langle A \rangle=\Tr(A\varrho)$ and due to (iv) we obtain:
\begin{equation} \label{mvps}
\langle A \rangle= \int _{\cal P}\psF_{A}^{(\ord)}(\pspoint)\W
^{(-\ord)}(\pspoint)\dif\mu(\pspoint).
\end{equation}
Therefore, the expectation value of an observable $A$ can be computed
as the average on $\cal P$ of $W^{(\sigma)}_A$ weighted by
$W^{(-\sigma)}_\varrho$. However, $W^{(\sigma)}_\varrho$ may be
negative at some points (\cite{hiletal84}), so that in general it does
not define a true probability distribution. For this reason
$W^{(\sigma)}_\varrho$ is sometimes called a {\it quasiprobability
distribution}.

Finally, the covariance property (v) means that the SW correspondence
commutes with the action of $G$, thus preserving the symmetry.



Notice that under the SW correspondence, an operator is mapped to a
set of functions labeled by $\sigma$, which is related to the operator
ordering prescription. In particular, $W^{(\sigma)}_A$ for $\ord = 0,
1$ and $-1$ are known, respectively, as symmetric, normal and
antinormal functions. The nomenclature is due to the fact that, for
the spinless particle, $\ord=0$ corresponds to the symmetric ordering
(\cite{moy49}), whereas $\ord = 1,-1$ are associated to the normal and
antinormal orderings, respectively (\cite{zhafengil90}).


\subsection{The SW kernel}

The vector space ${\cal O}({\cal H})$ equipped with the bilinear form
$(\opA,\opB)=\Tr(\opA^\dagger \opB)$ is a Hilbert space and due to the
Riesz theorem, there exists a $\sigma$-parameterized set of
operator-valued functions on $\cal P$, $\Delta^{(\sigma)}\in
C^\infty({\cal P})\otimes{\cal O}({\cal H})$, called the SW kernels,
such that:
\begin{equation} \label{sws}
\psF_A^{(\ord)}(\pspoint)=\Tr(\hA\sws ^{(\ord)}(\pspoint)).
\end{equation}

The properties (ii)-(v) introduced above are equivalent to the
following conditions on the kernels (\cite{vargra89, briman98,
briman99}):
\begin{enumerate}
\item[(ii$'$)]$$\Delta^{(\sigma)}(\pspoint)=
\Delta^{(\sigma)}(\pspoint)^\dagger$$
\item[(iii$'$)]$$\int_{\cal P}\Delta^{(\sigma)}(\pspoint)\dif\mu(\pspoint)=1$$
\item[(iv$'$)]$$\Delta^{(\sigma)}(\pspoint)=\int_{\cal
P}\Tr\Big(\Delta^{(\sigma)}(\pspoint)\Delta^{(-\sigma)}(\pspoint')\Big)
\Delta^{(\sigma)}(\pspoint') \dif\mu(\pspoint')$$
\item[(v$'$)]$$\Delta^{(\sigma)}(g\cdot\pspoint)=
\Delta^{(\sigma)}(\pspoint)^{g^{-1}}, \ \forall g\in G.$$

\end{enumerate}

Remarkably, the tracial property (iv$'$) allows to invert the SW map
(\ref{sws}), yielding a generalization of the Weyl rule:
\begin{equation}\label{inverseSW}
\hA=\int _{\cal
P}\psF_{A}^{(\ord)}(\pspoint)\sws^{(-\ord)}(\pspoint)\dif\mu(\pspoint).
\end{equation}
\noindent Notice that the kernel involved in (\ref{sws}) is
$\Delta^{(\sigma)}$, whereas the kernel entering formula
(\ref{inverseSW}) is $\Delta^{(-\sigma)}$. The case $\sigma=0$ is
privileged and $\Delta^{(0)}$ is said to be a self-dual kernel.

\subsection{The star product}

Once the Stratonovich-Weyl correspondence has been constructed, the
connection with the notion of $\star$-product is straightforward. The
$\star$-product is obtained by transferring to $\cal P$ the
associative algebra structure of ${\cal O}({\cal H})$ through the SW
map. That is,
\begin{equation}\label{defstarproduct}
(\psF_{\opA}^{(\ord)}\star
\psF_{\opB}^{(\ord)})(\pspoint):=\psF_{\opA\opB}^{(\ord)}(\pspoint).
\end{equation}
for any two operators $\opA$ and $\opB$\footnote{A more general
relation mixing different orderings can be given (see for example
\cite{briman99}), but we will not discuss it in this work.}.

At this point, a natural question arises. Given an arbitrary
$\star$-product on $C^\infty(\cal P)$, is there a Poisson bracket
associated to it in a canonical way? The answer is positive and the
proof is easy. Using associativity, a direct computation shows that
\begin{equation}
\{f,g\}=\frac{f\star g-g\star f}{\hbar}\quad \mbox{mod}\ \hbar,\quad
f,g\in C^\infty(\cal P)
\end{equation}
is a Poisson bracket. This is a nice way to see that any quantum
system has a corresponding classical system which is obtained in the
limit $\hbar\rightarrow 0$ by replacing commutators by Poisson
brackets. For example, we can write the transformation of the
von Neumann equation to phase space:
\begin{equation}
\fl
\qquad
\qquad
\partial_t \varrho
=
-\frac{\iu}{\hbar} [H, \varrho]
\qquad
\longmapsto
\qquad
\partial_t \W ^{(\ord)}
=
-\frac{\iu}{\hbar} ( \psF_{H}^{(\ord)} \star \W ^{(\ord)} - \W ^{(\ord)}
 \star \psF_{H}^{(\ord)})
\end{equation}
where $H$ is the Hamiltonian of the system. Hence, the star product
determines the dynamics on the phase space and reproduces the
classical Hamilton equations in the limit $\hbar\rightarrow 0$.


\subsection {An example: the spinless non-relativistic particle}

The classical phase space of the spinless non-relativistic particle is
${\mathbb R}^2$ with canonical Poisson bracket in coordinates $(q,p)$:
\begin{equation}
\{q,p\}=1.
\end{equation}

The Hilbert space of the quantum system is $L^2({\mathbb R})$ and its
dynamical symmetry group is the Heisenberg-Weyl group $H_3$. The Lie
algebra of $H_3$ is generated by three elements $I,a$ and $a^\dagger$
with Lie brackets:
\begin{equation}
[a,I]=[a^\dagger,I]=0,\quad [a,a^\dagger]=I.
\end{equation}
i.e. $a^\dagger$ and $a$ are creation and annihilation bosonic
operators, respectively.

The SW correspondence in this case maps each operator $A(\hat q, \hat
p)$ into $C^\infty({\mathbb R}^2)$, where
\begin{equation}
(\hat q\psi)(q_0):=q_0\psi(q_0),\quad (\hat
p\psi)(q_0):=-\iu\hbar\frac{\partial\psi}{\partial q}\vert_{q_0}, \quad
\forall q_0\in {\mathbb R}
\end{equation}
for any $\psi\in L^2({\mathbb R})$.

Using the standard complex
coordinate ${z} = (q + \iu p)/\sqrt{2\hbar}$ the SW kernel reads:
\begin{equation}\label{kernelexample}
\sws^{(\sigma)}({z}) =1/\pi\int _{\mathbb C}
e^{\frac{\sigma}{2}|\xi|^2} e^{\xi^* {z} - \xi {z}^*} e^{ \xi a^{+} -
\xi^{*} a}\dif^2 \xi.
\end{equation}

Let us focus on the case $\sigma=0$ (symmetric case). For this
particular value of the parameter $\sigma$, (\ref{kernelexample})
becomes the famous Wigner function \cite{wig32}:
\begin{eqnarray}
W_ {A}^{(0)}(q,p) = \int_{-\infty}^{+\infty} \e^{\iu p u/\hbar} \langle q -
\case{1}{2} u \vl A \vl q + \case{1}{2}u \rangle \dif u.
\end{eqnarray}

The $\star$-product is the Moyal product, whose closed form was
 introduced by Groenewold in the forties \cite{gro46}:
\begin{eqnarray}
\psF_{\opA\opB}^{(0)} (q,p)
=
\psF_{\opA}^{(0)} 
\e^{-\iu\hbar\Gamma/2}
\psF_{\opB}^{(0)}\vert_{(q,p)},
\qquad
\quad
\Gamma 
:=
\frac{\overleftarrow {\partial}} {\partial p}
\frac{\overrightarrow {\partial}}{\partial q}
-
\frac{\overleftarrow {\partial}} {\partial q}
\frac{\overrightarrow {\partial}}{\partial p}.
\end{eqnarray}

A useful representation of the star product was introduced by Bopp
\cite{bop56}. He found that, for any two operators $\opA(\hat q,\hat p)$
and $\opB(\hat q,\hat p)$, the $\star$-product can be casted in the
following form:
\begin{eqnarray}
\nonumber
\psF_{\opA}^{(0)}
\star
\psF_{\opB}^{(0)}
=
\opA(\bopQ, \bopP) \psF_{\opB}^{(0)}
\\
\label{bopppar}
\bopQ := q +\frac {\iu\hbar}{2} \frac {\partial}{\partial p}
,
\quad
\bopP := p -\frac {\iu\hbar}{2} \frac {\partial}{\partial q}
\end{eqnarray}
where now $\opA (\bopQ,\bopP)$ should be understood as an operator acting on
$C^\infty({\cal P})$. The operators $\bopQ$ and $\bopP$ are usually known as
{\it Bopp operators}. Observe that $[\bopQ,\bopP] = \iu\hbar$, which can be
checked directly from (\ref{bopppar}). As a consequence, $\bopQ$ and $\bopP$
are operators on $C^\infty({\cal P})$ satisfying the canonical
commutation relations.

Using the reality condition of the SW kernel we can write the time
evolution of the symbol of the density matrix, $W_\varrho^{(0)}$, in a
simple way:
\begin{eqnarray}
\label{vnpsp}
\partial_t \W^{(0)}
=
\frac{2}{\hbar}
{\rm Im}
\Big (
 H(\bopQ,\bopP) \Big )  \W^{(0)}
\end{eqnarray}
%
with $H$ the Hamiltonian of the system. The classical limit is very
easily obtained now. Expanding up to first order in $\hbar$,
\begin{equation}
H(\bopQ,\bopP) = H(p,q) + \frac{\iu\hbar}{2}(\partial _q
H\frac{\partial}{\partial p} - \partial _p H\frac{\partial}{\partial
q})+O(\hbar^2)
\end{equation}
we recover the classical Liouville equation:
\begin{equation}
\partial _t\W^{(0)} = \{H, \W^{(0)}\}.
\end{equation}


\section{The spin case}\label{sec:spincase}

We now turn to spin systems, our main interest in this
work. The dynamical symmetry group is ${\rm SU(2)}$, and the phase
space is the unit sphere ${\rm S}^2$. The classical Poisson bracket is
suitably written in terms of the components of
\begin{equation}\label{eq:defofm}
\fl
\qquad
{\bf m} (\theta,\phi) = ( \sin \theta \cos \phi, \sin \theta \sin \phi,
\cos \theta).
\end{equation}
Namely\footnote{Summation over repeated indices must be understood
throughout the paper.},
\begin{equation}\label{poiss}
\{
m_i
, m_j
\}
=
\frac {1}{S}
\epsilon _{ijk} m_k,
\end{equation}
where $\epsilon_{ijk}$ is the Levi-Civita symbol. Strictly, the
Poisson bracket should be $\{ m_i , m_j \} = g \epsilon _{ijk} m_k$,
where $g$ is the {\it gyromagnetic ratio}, which is related to the
quantum spin number by $S= \mu_B/g\hbar$, with $\mu_B$ is the magnetic
moment. From now on we set $\hbar = \mu_B =1$, so that $S = 1/g$. The
classical Liouville equation may be expressed (\cite{gar2000}) as
%
%
\begin{equation}
\label{liouvilles}
\hspace{-1cm}\partial_t\W^{(\ord)} =
\{H, \W^{(\ord)} \}
=
-
\frac{1}{S}
\frac {\partial}{\partial {\bf m}}
\cdot
\Big (
{\bf m}
\times
{\bf B}_{\rm eff}
\Big )
\W^{(\ord)}
,
\qquad
{\bf B}_{\rm {\rm eff}} := - \frac{\partial H}{\partial {\bf m}}
\; ,
\end{equation}
%
with the divergence $(\partial / \partial {\bf m} )\cdot {\bf A}  = \sum_i
\partial A_i /\partial m_i$.
The purpose of this section is to generalize the Bopp operators for
the spin case, which simplifies the study of the dynamics in
phase-space reported in the literature \cite{kli02}.

\subsection {The SW kernel and the star product}

The SW kernel reads \cite{aga81, vargra89, briman98, briman99}:
\begin{equation}
\label{swsspins}
\sws^{(\ord)}(\theta, \phi)
=
\sqrt {\frac{4 \pi}{2S +1}}
\sum_{l=0}^{2S}
\langle S, S; l,0 \vl S,S \rangle ^{-\sigma}
\sum_{m=-l}^{l}
\T_{lm} Y_{lm}^{*} (\theta, \phi)
\end{equation}
where $\langle j, j; l,0 \vl j,j \rangle$ are  Clebsch-Gordan
coefficients, $\T_{lm} = \sqrt {(2l+1)/(2S+1)}
\times\sum_{j, j'} \langle S, j; S, j' \vl S, j'\rangle \vl S,
j'\rangle \langle S, j \vl$ are the irreducible tensor operators and
$Y_{lm}$ are the spherical harmonics.

Recently, Klimov and Spinoza derived a differential form for the star
product in the spin case \cite{kliesp02}:
\begin{equation}
\label{ksp}
\nonumber
\fl
\
\psF_{\opA}^{(\ord)}
\star
\psF_{\opB}^{(\ord)}
=
N_S
\sum _{j}
a_j
\widetilde {F}^{\ord-1} (\cas^2)
\Big[\Big(
\opS^{+ (j)}
\widetilde {F}^{1-\ord} (\cas^2)
\psF_{\opA}^{(\sigma)}\Big)
\otimes
\Big(
\opS^{- (j)}
\widetilde {F}^{1-\ord} (\cas^2)
\psF_{\opB}^{(\sigma)}\Big)
\Big]
\end{equation}
where $\cas^2$ is the Casimir operator on the sphere,
\begin{equation}
\fl
\qquad
\cas ^2
= 
-
\left [
\frac {\partial ^2}{\partial \theta^2}
+
\cot \theta
\frac {\partial}{\partial \theta}
+
\frac {1}{\sin ^2 \theta}
\frac {\partial ^2}{\partial \phi^2}
\right ]
,
\qquad
\qquad
\cas ^2 Y_{lm}
=
l (l+1) Y_{lm}
,
\end{equation}
\begin{equation}
\fl
\qquad
\widetilde {F} (\cas^2) Y_{lm}
= F (l) Y_{lm}
,
\qquad
\qquad
\qquad
\qquad
F(l) =
\sqrt {(2S + l +1)!(2S- l)!}
,
\end{equation}
\begin{equation}
\label{opS}
\fl
\qquad
\opS ^{\pm (j)}
=
\prod_{k=0}^{j-1}
\left (
k \cot \theta
- 
\frac {\partial}{\partial \theta}
\mp
\frac {\iu}{\sin \theta}
\frac {\partial}{\partial \phi}
\right )
,
\qquad
a_j
=
\frac {(-1)^j}{j! (2S + j +1)!},
\end{equation}
and $N_S = \sqrt {2S +1} F^\ord(0)$.


\subsection {Bopp operators in the spin case}\label{sec:boppspins}

The evaluation of (\ref{ksp}) seems quite involved. Here, we show that
it can be simplified by generalizing the Bopp operators
(\ref{bopppar}) for the spin case. That is, we look for operators $\bopps_i^{(\ord)}$ acting on $C^\infty(\cal P)$ such that:
\begin{equation}
\label{boppgral}
\psF_{S_i}^{(\ord)} \star \psF_{A}^{(\ord)} = \bopps_i^{(\ord)}
\psF_{A}^{(\sigma)},\quad\quad i=1,2,3.
\end{equation}

Let us invoke the expression derived in the Appendix for
$\psF_{S_3}^{(\ord)} \star Y_{lm}$, Eq.  (\ref{boppszap}). Noticing
that $m_3 = \cos \theta$, $ L_3 = -i \partial_{\phi} $, $ \iu ({\bf m
\times L})_3 = \sin \theta \partial_\theta $ and recalling that every
function on the sphere is a linear combination of spherical harmonics
we deduce that for any operator $A$:
\begin{equation}
\psF_{S_3}^{(\ord)} \star \psF_{A}^{(\ord)}
=
\left [
m_3
\tcoef_1^{(\ord)} (\cas^2)
+
\iu ({\bf m} \times {\bf L})_{3}
\tcoef_2^{(\ord)} (\cas^2)
+
\frac{1}{2}
L_3
\right ]
\psF_{A}^{(\ord)}.
\end{equation}
where $\tcoef_i^{(\sigma)}
(\cas^2)Y_{lm}={\coef_i^{(\ord)}}(l)Y_{lm},\ i=1,2$ and
\begin{eqnarray}
\nonumber
\fl
\coef _1 ^{(\ord)}(l) &=&
\frac {F^{1-\sigma} (l)} {2 (2l +1)}
\Big [
F^{\sigma-1}(l+1) 
\left (2S+ l+2 \right )
(l+1)
-
F^{\sigma-1}(l-1) 
\left (l-2S-1 \right )
l
\Big ]
\\
\label{fs}
\fl
\coef_2 ^{(\ord)}(l) &=&
\frac {F^{1-\sigma} (l)} {2 (2l +1)}
\Big [
F^{\sigma-1}(l+1) 
\left (2S+ l+2 \right )
+
F^{\sigma-1}(l-1) 
\left (l-2S-1 \right )
\Big ].
\end{eqnarray}
The vector {\bf m} has been defined in (\ref{eq:defofm}) and ${\bf L}$
is the angular momentum operator acting on the sphere. Explicitly,
\begin{equation}\label{commL}
{\bf L} = -\iu \left (
{\bf m} \times \frac {\partial}{\partial {\bf m}}
\right )
\end{equation}
which satisfies $[L_i, L_j] = \iu \epsilon_{ijk} L_k$. The following
identities are useful:
\begin{eqnarray}
\label{conmvec}
[m_i, L_j] = \iu \epsilon_{ijk} m_k
;
\qquad
[({\bf m \times L})_i, L_j] = \iu \epsilon _{ijk} ({\bf m \times L})_k
\; .
\end{eqnarray}
%
Hence,
\begin{equation}
\bopps _3^{(\ord)} = 
m_3\tcoef_1^{(\ord)}(\cas^2) + \iu
({\bf m} \times {\bf L})_{3} \tcoef_2^{(\ord)}(\cas^2) + 
\frac{1}{2} L_{3}.
\end{equation}

The covariance of the star product at the infinitesimal level implies
that
%
\begin{equation}
\label{cov}
 \psF_{ \iu \epsilon_{ijk} S_k A}^{(\ord)} = \psF_{[S_i, S_j] A}^{(\ord)}=
[L_i, \bopps_j^{(\ord)}] \psF_A^{(\ord)}
\end{equation}
%
whence
\begin{equation}
[\bopps_i^{(\ord)}, L_j] = \iu \epsilon _{ijk} \bopps _k^{(\ord)}.
\end{equation}
Using Eq.  (\ref{cov}) along with (\ref{commL}) and (\ref{conmvec}) we
obtain:
\begin{equation}
\bopps _i^{(\ord)} = 
m_i \; \tcoef_1^{(\ord)}(\cas^2) + \iu
({\bf m} \times {\bf L})_{i} \; \tcoef_2^{(\ord)}(\cas^2) + 
\frac{1}{2} L_{i}
\end{equation}

The operators ${\cal S}_i^{(\sigma)}$ are the sought generalization of
the Bopp operators to the spin case. Now, by the associativity of the
star product, we can extend (\ref{boppgral}) to any function
$\opA(S_1,S_2,S_3)$ of the spin operators:
\begin{equation}
\label{mrb}
\psF_{\opA\opB}^{(\ord)}
=
\opA (\bopps _1^{(\ord)}, \bopps_2^{(\ord)}, \bopps _3^{(\ord)}) 
\psF_{\opB}^{(\ord)}
\end{equation}
%
where, equivalently to the spinless particle case (see
Eq.  (\ref{bopppar})), $\opA (\bopps _1^{(\ord)}, \bopps_2^{(\ord)},
\bopps_3^{(\ord)})$ is the same function of its arguments as
$\opA(S_1,S_2,S_3)$, now acting on $C^\infty({\cal P})$. In order to
ease the notation we will simply write $\opA(\bopps ^{(\ord)})$
instead of $\opA (\bopps_1^{(\ord)}, \bopps_2^{(\ord)}, \bopps
_3^{(\ord)})$.

The von Neumann dynamics is governed by the equation (compare with
(\ref{vnpsp})):
%
\begin{equation}
\label{vnpss}
\partial_t \W ^{(\ord)}
=
2
{\rm Im}
\Big (
 H(\bopps^{(\ord)}) 
\Big )
 \W ^{(\ord)}
\end{equation}
%
Besides $[\bopps _i^{(\ord)}, \bopps _j^{(\ord)}] = \iu \epsilon_{ijk}
\bopps _k^{(\ord)}$.
In this sense Bopp operators are the phase-space analogues of the
corresponding Hilbert-space operators.

\vskip0.5 cm
{\it Remark:}
\vskip0.1cm

In the $\ord = 0, \pm 1$ cases we can write:
%
\numparts
\begin{eqnarray}
\fl
\coef_1^{(\ord)} =
\left \{
\begin{array}{cc}
S & \ord = -1
\\
\frac{1}{2 (2l+1)}
\left [
(l+1)
\sqrt {(2S+1)^2 - (l+1)^2}
+
l
\sqrt {(2S+1)^2 + l^2}
\right ] & \ord = 0
\\
S+1 & \ord = 1
\end{array} 
\right .
\\
\fl
\coef_2^{(\ord)} =
\left \{
\begin{array}{cc}
-\frac {1}{2} & \ord = -1
\\
\frac{1}{2 (2l+1)}
\left [
\sqrt {(2S+1)^2 - (l+1)^2}
-
\sqrt {(2S+1)^2 + l^2}
\right ] & \ord = 0
\\
\frac{1}{2} & \ord = 1
\end{array} 
\right .
\end{eqnarray}
\endnumparts
%
The form of $\bopps^{(\pm 1)}_i$ was already found by Takahashi and
Shibata in \cite{takshi76} by extending the construction of Cahill and
Glauber. We have recovered their result and generalized it to any
value of $\ord$ in terms of the symbol (\ref{sws}) for the
differential form of the star product of Klimov and Espinoza
(\ref{ksp}).
This covers the important case of symmetric ordering.


\subsection {Large S}

We turn to the large S (or classical) limit. Firstly, notice that for
$S \gg 1$ we can write \cite{kliesp02}:
\begin{equation}
 {F} (l)
=
\frac {(2S+1)!}{\sqrt {2S+1}}
\left [
1
+
\frac{1}{2 (2S+1)}
l(l+1)
+
\Or \left ( \frac {1}{S^2} \right)
\right ].
\end{equation}
Inserting the above expansion in (\ref{fs}) we inmediately find that:
\begin{eqnarray}
\label{asympfs}
&&2\tcoef_1^{(\sigma)} = 2S+1 + \ord + \frac{1}{2(2S+1)} (\cas
^2 +1) (\ord^2 -1)+ \Or \left (\frac {1}{S ^2} \right ),\cr
&&2\tcoef_2^{(\sigma)} = \ord + \frac {1}{2 (2S+1)} \big (
\ord^2 -1 \big ) + \Or\left (\frac {1}{S ^2} \right ).
\end{eqnarray}
The above expressions allow us to evaluate the quasiclassical
evolution of $\W^{(\sigma)}$ in a simple way (see also
\cite{kliesp05}). This is an alternative to the spin coherent path
integral formalism used in \cite{StoParGar00},
\cite{GarKocParSto03}. In the classical limit, $\tcoef _1^{(\sigma)} =
S + (\ord+1)/2 + \Or(1/S)$ and $\tcoef _2^{(\sigma)} =\ord/2 + \Or
(1/S)$. Let $H(S_1,S_2,S_3)$ be the Hamiltonian. Also in the classical
limit,
\begin{equation}
\label{clherm}
{\rm Im} \Big( H (\bopps^{(\ord)}) \Big)
=
\frac {-\iu}{2 S} 
\frac {\partial H}{\partial  m_i}  L_i
\end{equation}
and using that for any ${\bf b}\in{\mathbb R}^3$,
\begin{equation}\label{eq:identity}
-\iu {\bf b} \cdot {\bf L} = \frac{\partial}{\partial{\bf m}} \cdot [
  {\bf m} \times {\bf b}].
\end{equation}
we recover the classical Poisson bracket (\ref{liouvilles}), obviously
independent of $\ord$.


\subsection {Quadratic Hamiltonians}\label{subsec:QuadHam}

We apply now the formalism to the following case:
%
\begin{equation}
H
=
-
D_{ij}
S_iS_j
-
B_i S_i,\hspace{2cm} D_{ij}=D_{ji}
\end{equation}
%
which is the most general Hamiltonian quadratic in the variables
$S_i$. This example covers many applications in magnetism \cite{white}
and quantum optics (see, e.g., \cite{agarpursin97,molsor99}).
With the  help of (\ref{mrb}) and (\ref{boppgral}) we obtain that
$
{\rm Im}(H)
=
- \iu\left[ 
D_{jk} L_j \left( m_k \tcoef_1^{(\sigma)}  + \iu
({\bf m} \times {\bf L})_k \tcoef^{(\sigma)}_2 \right)+
B_j L_j/2
\right]
$
and using (\ref{eq:identity}) we finally get:
\begin{equation}
\label{quadratic}
\partial_t \W^{(\sigma)}
=
-\frac{1}{S}\frac {\partial}{\partial {\bf m}} \cdot \Big [
{\bf m} \times {\bf B}_{\rm eff}^Q
\Big ] \W^{(\sigma)}
\end{equation}
with
\begin{equation}
\label{BeffQ}
[{\bf B}_{\rm eff}^Q]_j
=
[{\bf B}_{\rm eff}]_j
+
2S D_{jk}
\Big [
m_k (
\tcoef_1^{(\sigma)}
-S)
+
\iu
({\bf m} \times {\bf L} )_k
\tcoef_2^{(\sigma)}
\Big ]
\end{equation}
Equation (\ref{quadratic}) is a compact way of writing the quantum
dynamics in phase-space. ${\bf B}_{\rm eff}^Q$ is defined in such a
way that the first term gives the classical effective field $[{\bf
B}_{\rm eff}]_j = S B_j + 2S^2 D_{jk} m_k$ (see Eq.
(\ref{liouvilles})) and the second one the quantum contributions. When
$D_{jk} =0$ the quantum time evolution equation (\ref{quadratic})
coincides with the classical one (\cite[Theorem 5]{vargra89}).  This
is analogous to the harmonic oscillator for the spinless particle
\cite{hiletal84} where the classical and quantum equation of motion in
phase space are the same. Notice, however, that although the
phase-space time evolution equations in the classical and quantum
cases are formally identical, the allowed solutions for the
differential equations are different at the classical and quantum
levels.

\vskip 0.5 cm
{\it Remark:}
\vskip0.1cm

Consider the particular case $H = -D S_3^2$. Then,
Eq. (\ref{quadratic}) simply reads:
\begin{equation}
\partial_t \W^{(\sigma)}
=
-2 D
\Big (
{\cos \theta
\tcoef} _1^{(\sigma)}
+
\sin \theta
\frac {\partial}{\partial \theta}
\tcoef^{(\ord)}_2
\Big )
\frac {\partial\W^{(\sigma)}}{\partial \phi}
\end{equation}
with ${\tcoef} _1^{(\sigma)}$ and ${\tcoef}_2^{(\sigma)}$ given in
(\ref{fs}). For $\ord = \pm 1$ we recover the results found by Klimov
(see equations (65) and (67) of Ref. \cite{kli02}). However, we obtain
a different result for $\ord = 0$ (Eq.  (66) in Ref. \cite{kli02}).

\vskip0.5cm

Finally, using the asymptotic forms of $\tcoef_1^{(\ord)}$ and
 $\tcoef _2^{(\ord)}$, Eq. (\ref{asympfs}), we obtain the
 quasiclassical form of (\ref{quadratic}). In the case $H = -D S_3^2$
 the result agrees with derivations previously appeared in the
 literature (see the Conclusions of Ref. \cite{kliesp05}).


\section {Application to Quantum Master Equations}\label{sec:quantumME}

The field of open quantum systems deals with the unavoidable
interaction between the system of interest and the environment
\cite{weiss}. Under assumptions such as weak coupling between the
environment and the system and fast bath dynamics \cite{mohsmi80,
calleg83}, the environment may be modelled as a collection of
oscillators. With these hypotheses one can write the total Hamiltonian
(system plus environment) as:
%
\begin{equation}
\label{clmodel}
\HTOT
=
\HS
+
\sum _\bindex
\frac{1}{2}
\left \{
P_\alpha^2
+
\omega^2 _\alpha
\left [
Q_\alpha 
+
\frac {c_\alpha}{\omega^2 _\alpha}
F
\right ]^2
\right \}
\end{equation}
where $\alpha$ is an oscillator index and the coupling terms $F$ are
functions of the system variables.

Classically the dynamics of the system degrees of freedom is
  formulated in terms of {\it Langevin} or {\it Fokker-Planck}
  equations \cite{zwanzig, risken}. At the quantum level, $Q_\alpha$ and
  $P_\alpha$ are the position and momentum operators and $F$ is a
  hermitian operator which depends on the system variables.
The total system is unlikely to be in a pure state and a
density-matrix description is required.
For observables depending only on the system variables, the required
object is the {\em reduced\/} density operator $\varrho=\Tr_{\rm
bath}(\varrho_{\rm tot})$, where one traces the bath variables out.
For {\em weak\/} system-bath coupling a closed dynamical equation for
$\varrho$ can be obtained by perturbation theory. This is the case of
many problems in quantum optics, chemical physics or magnetism
\cite{weiss}. As far as the classical limit is concerned, a
semiclassical quantum master equation would be sufficient. Let $T$ be
the temperature of the bath and $\gamma$ the damping coefficient
(measuring the coupling strength). Considering that $\gamma /(ST) \ll
1$ one arrives to \cite{calleg83pa, brepet}:
%
\begin{equation}
\label{dme}
\partial_t \varrho
=
-\iu [\HS, \varrho]
-
\gamma T
\Big (
[\F, \F \varrho] - \frac{1}{2T} [\F, [\HS, \F]\varrho] + {\rm h.c.}
\Big )
\end{equation}
where h.c. means ``hermitian conjugate''. The term $-\iu [\HS, \varrho]$ gives
the isolated-spin unitary evolution and the rest encodes the bath influence.

Now, we address the problem of working out the 
phase-space transform of
(\ref{dme}).
In the spinless case for the generic Hamiltonian $\HS = p^2/2m + V(q)$
and bilinear coupling, $\F = q$, the master equation (\ref{dme}) is
the celebrated {\it Caldeira-Leggett} equation \cite{calleg83pa}.
Then the phase-space transformation gives the first quantum
corrections to the Klein-Kramers equation, which is nothing but the
Fokker-Planck equation for a particle in a potential \cite{risken}.

In order to obtain the quantum corrections to the classical
Fokker-Planck equations for spins, we will make use of the generalized
Bopp operators. Using the results of Section \ref{sec:boppspins} for
transforming Hilbert space operators into operators acting on
$C^\infty({\cal P})$ we obtain:
\begin{equation}
[\F, \F \varrho] + {\rm h.c.} 
\quad
\mapsto
\quad
\Big (
\F (\bopps^{(\ord)}) - {\rm c.c.}
\Big )^2 \W^{(\ord)}
\end{equation}
and
\begin{equation}
\hspace{-1.5cm}[\F, [\HS, \F]\varrho]
+
{\rm h.c.}
\quad
\mapsto
\quad
\Big (
\F (\bopps^{(\ord)}) - {\rm c.c.}
\Big )
\Big (
[\HS (\bopps^{(\ord)}), \F(\bopps^{(\ord)})]
- {\rm c.c.}
\Big )
\W^{(\ord)}
\end{equation}
where c.c. means ``complex conjugate''. Then, Eq.  (\ref{dme}) reads in
the phase-space formulation:
\begin{eqnarray}
\label{qfp}
\fl
\partial_t \W^{(\ord)}
=
&\Bigg [&
2{\rm Im}
\big (
\HS (\bopps^{(\ord)})
\big )+
\\
&+& 4\gamma T
\bigg [
{\rm Im}^2
\big (
F (\bopps^{(\ord)})
\big ) \nonumber 
-
\frac {1}{2T}
 {\rm Im}
\big (
F (\bopps ^{(\ord)})
\big )
 {\rm Im}
\big (
[\HS (\bopps^{(\ord)}), \F (\bopps^{(\ord)})]
\big )
\bigg ]
\Bigg ]
\W^{(\sigma)}
\end{eqnarray}
This is the phase-space equivalent of the master equation (\ref{dme})
and should be viewed as the quantum generalization of the classical
Fokker-Planck equations. We emphasize that the transformation to the
phase-space formalism has been rather simple thanks to (\ref{mrb}).

\subsection{The open isotropic spin versus the open harmonic oscillator}

To gain some insight we particularize Eq. (\ref{qfp}) to an specific example.
Our intention here is to highlight the differences between the spinless
particle and the spin system.
For that, we closely follow the work of Caldeira and Leggett on the
quantum master equation for a spinless particle \cite{calleg83pa}.
In this case the phase-space transform of the last two terms of
(\ref{dme}) gives the dissipation and diffusion terms of the classical
Fokker-Planck equation \cite{calleg83}.  In particular, for the
harmonic oscillator, both the {\it Caldeira-Leggett} and {\it
Fokker-Planck} equations are identical (see the discussion of Section
\ref{subsec:QuadHam}).

Similarly, in the spin case we also choose a bilineal coupling, $\F=
\xi_j S_j$, with $\xi_j$ some real constants.
For spin systems $[\HS, \F]$ is in general $\HS$-dependent and a
generic form of the time evolution equation cannot be given.
For the sake of simplicity we take $\HS = - B_i S_i$, i.e. the
isotropic spin ($D_{ij} = 0$ in (\ref{quadratic})). This is the
closest analogue to the open quantum oscillator discussed above
(section \ref{subsec:QuadHam}). Then, the quantum master equation
(\ref{dme}) follows in phase-space from (\ref{qfp}):
\begin{eqnarray}
\label{qfpbil}
\hspace{-2.5cm}
\partial _t \W^{(\ord)}
=
-
\frac{1}{S}
\frac {\partial}{\partial {\bf m}}
\cdot
\Bigg \{
({\bf m} \times {\bf B}_{\rm eff})
&-&
{\bf m}
\times
\tL
\Big [
{\bf m}
\times
\Big (
{\bf B}_{\rm eff}
-
T
\frac{\partial}{\partial {\bf m}}
\Big )
+
{\bf M}
\times
{\bf B}_{\rm eff}
\Big ]
\Bigg \}
\W^{(\ord)}
\end{eqnarray}
where $[{\bf B}_{\rm eff}]_j = S B_j$ is the classical effective field
[see Eq. (\ref{liouvilles})], $\tL$ has components
\begin{equation}
\label{tL}
\tL_{jk}
=
\frac{\gamma}{S}
\frac  {\partial F}{\partial m_j}
\frac  {\partial F}{\partial m_k}
\; ,
\end{equation}
(if $\F$ is linear, then $\tL_{jk} = S \gamma \xi_j \xi_j$) and ${\bf
M} = {\bf m}({\tcoef}_1^{(\sigma)} - S)+\iu ({\bf m}\times {\bf L})
\tcoef_2^{(\sigma)}$. Observe that ${\bf M} = \Or(1/S)$, so the last
term vanishes as $S \to \infty$ recovering the classical Fokker-Planck
equation \cite{gar2000}.

Hence, even in the case where both system ($\HS$) and coupling ($\F$)
are linear in the spin variables, classical and quantum Fokker-Planck
equations are different, unlike the open harmonic oscillator.
Within the phase-space formalism this difference can be worked out
explicitly.

\subsection {The classical limit of (\ref{qfp}):  general case}

To finish, let us go back to (\ref{qfp}) and compute its classical
limit for general $\HS$ and $\F$.
Using (\ref{clherm}) one finds that,
\begin{equation}
{\rm Im}^2
\bigg (
F (\bopps^{(\ord)})
\bigg )
\stackrel {S \to \infty} \longrightarrow
\frac {1}{4S^2}
L_j 
\frac  {\partial F}{\partial m_j}
\frac  {\partial F}{\partial m_k}
L_k 
\end{equation}
%
\begin{equation}
{\rm Im} \big ( [\HS (\bopps^{(\ord)}), \F (\bopps^{(\ord)})] \big )
\stackrel {S \to \infty}\longrightarrow -1/S \{\HS, \F \}
\end{equation}
Hence,
\begin{equation}
 {\rm Im} \bigg ( F (\bopps^{(\ord)}) \bigg ) {\rm Im} \bigg ( [\HS
(\bopps^{(\ord)}), \F (\bopps^{(\ord)})] \bigg ) \stackrel {S \to
\infty} \longrightarrow \frac {\iu}{2S^2} L_j \frac {\partial
F}{\partial m_j} \{\HS, \F \}
\end{equation}
%
Since $\{\HS, \F \} = \frac{1}{S}
\epsilon_{ijk}
(\partial_{m_j} \HS)( \partial_{m_k} F) m_i
$
we finally obtain:
\begin{equation} \label{clfp}
\hspace{-1cm}\partial_t \W^{(\ord)}
=
-
\frac{1}{S}
\frac {\partial}{\partial {\bf m}}
\cdot
\Bigg \{
({\bf m} \times {\bf B}_{\rm eff})
-
{\bf m}
\times
\tL
\Big [
{\bf m}
\times
\Big (
{\bf B}_{\rm eff}
-
T
\frac{\partial}{\partial {\bf m}}
\Big )
\Big ]
\Bigg \}
\W^{(\ord)}
\end{equation}
%
where $\tL_{jk}$ has been defined in (\ref{tL}) and ${\bf B}_{\rm
eff}$ in (\ref{liouvilles}). Eq.  (\ref{clfp}) is nothing but the
classical Fokker-Planck equation for classical spins
[cf. Eq. (\ref{qfpbil})] \cite{garishpan90} (see also \cite{gar2000}).


\section {Conclusions}

The phase-space formulation of Quantum Mechanics provides  deep
insight into the quantum-classical correspondence, mainly due to the
fact that the mathematical nature of the observables does not change
when going from the classical to the quantum theory: they are always
functions on the phase space.

It is well-known that for non-relativistic spinless particles Bopp
operators simplify the manipulations in the phase space formalism
\cite{bop56, hiletal84}. In this paper we have generalized Bopp
operators for spin systems and applied them to the problem of the
classical limit of spins as well as to open quantum spin systems.

The dynamics of open quantum systems has received a renewed attention
 due to its role in explaining the emergence of the classical world
 from quantum mechanics \cite{zur91}. Open quantum systems are handled
 with equations for the reduced density matrix,
 cf. Eq. (\ref{dme}). They consist of the unitary evolution (von
 Neumann) and a non-unitary term which incorporates the bath
 influence. For the spinless case the transformation of the time
 evolution equations into phase space yields the quantum
 generalization of the corresponding Fokker-Planck equations. We have
 accomplished this task in the spin case with the help of the
 generalized Bopp operators. We have worked out the classical limit
 ($S \to \infty$) of the quantum master equation for the density
 matrix, and recovered the classical Fokker-Planck equations.

\ack

We acknowledge J.~L.\ Garc\'{\i}a-Palacios and
J.~M. Gracia-Bond\'{\i}a for useful discussions. D. Z. acknowledges
financial support through DGA project {\tt PRONANOMAG} and grant no.\
B059/2003.


\appendix

\section{}

We present here the details for the calculation of Eq.  (\ref{boppgral}).
For that we compute explicitely $\psF_{S_3}^{(\ord)} \star Y_{lm}$, using
(\ref{ksp}).
We first notice that \cite{kliesp02}:
\begin{equation}
\psF_{S_3}^{(\sigma)}
=
\left (
\frac {S}{S+1}
\right )^{-\ord/2}
\sqrt {S (S+1)}
\cos \theta
\sim
Y_{1,0}
\end{equation}
Using that $\widetilde {F} (\cas ^2) \cos \theta = F(1) \cos \theta$ and that in the sum of
(\ref{ksp}) $j_{\rm max} = 1$, together with
\begin{equation}
a_0 = \frac{1}{(2S +1)!}
;
\qquad
a_1 = - \frac {a_0}{2S +2}
\end{equation}
we can write,
\begin{equation} \label{sp0}
\fl
\psF_{S_3}^{(\ord)} \star Y_{lm}
=
\frac{F^{1-\ord}(l)}{2}
\left(
\widetilde {F}^{\ord-1} \Big [ \cos \theta Y_{lm} \Big ]
-\frac{1}{2S +2} \widetilde {F}^{\ord-1}
\Big [
\Big (S^{+ (1)} \cos \theta \Big )
\Big (S^{-(1)} Y_{lm}
\Big )
\Big ]
\right)
\end{equation}
%
here $\opS^{+ (1)}$ acts only over $\cos \theta$,
and   $\opS^{- (1)}$ over $Y_{lm}$
with, see Eq. (\ref{opS}):
\begin{eqnarray}
\opS^{\pm (1)}
=
-
\frac {\partial }{\partial \theta}
\mp
\frac {\iu}{\sin \theta}
\frac {\partial}{\partial \phi}
\end{eqnarray}
Next we make use of the relations:
\begin{eqnarray}
\label{relations}
\nonumber
\cos \theta Y_{lm}
&=&
\alpha _1 Y_{l+1, m}
+
\alpha _2 Y_{l-1,m}
\\
\sin \theta \frac{\partial}{\partial \theta} Y_{lm}
&=&
\beta _1 Y_{l+1, m}
+
\beta _2 Y_{l-1,m}
\end{eqnarray}
with
\begin{eqnarray}\nonumber
&&\alpha _1 = \sqrt{\frac{(l -m +1)(l+m+1)}{(2l+1)(2l+3)}}\ ,\quad
\alpha_2 = \sqrt{\frac{(l-m)(l+m)}{(2l-1)(2l+1)}}\\[10pt]
&&\beta_1 =\alpha_1 l,\quad \beta_2=-\alpha_2 (l+1).
\end{eqnarray}
Manipulating the terms entering (\ref{sp0}) we find:
\begin{eqnarray}
\nonumber
F^{1-\ord} (l)\widetilde{F}^{\ord-1} \Big [ \cos \theta Y_{lm} \Big ]
&=&
\alpha _1 F^{1-\ord} (l) F^{\ord-1} (l+1) Y_{l+1 m}
\\
&+&
\alpha _2 F^{1-\ord}(l) F^{\ord-1}(l-1) Y_{l-1 m} 
\end{eqnarray}
and
\begin{eqnarray}
\fl
\qquad
F^{1-\ord}(l)
\widetilde {F}^{\ord-1}
\Big [
\Big (S^{+ (1)} \cos \theta \Big )
\Big (S^{-(1)} Y_{lm}
\Big )
\Big ]
&=&
-\beta _1  F^{1-\ord} (l) F^{\ord-1} (l+1) Y_{l+1 m}
\\ \nonumber
&-& \beta_2  F^{1-\ord}(l) F^{\ord-1}(l-1) Y_{l-1 m} 
\\ \nonumber
&+& \iu \frac {\partial}{\partial \phi}  Y_{l m}.
\end{eqnarray}
Now, taking into account that
\begin{eqnarray}
Y_{l-1m}
&=&
\frac{1}{(2l+1) \alpha_2}
\big (
l \cos \theta 
-
\sin \theta \frac {\partial}{\partial \theta}
\big ) Y_{lm}
\\
Y_{l+1m}
&=&
\frac{1}{(2l+1) \alpha_1}
\big (
(l+1) \cos \theta 
+
\sin \theta \frac {\partial}{\partial \theta}
\big ) Y_{lm}
\end{eqnarray}
we get:
\begin{eqnarray}
\label{KA1}
\nonumber
\fl
F^{1-\ord} (l)\widetilde{F}^{\ord-1} \Big [ \cos \theta Y_{lm} \Big ]
&=&
F^{1-\ord} (l) F^{\ord-1}(l+1)
\frac{1}{2l +1}
\left [  (l+1) \cos \theta + \sin \theta \frac{\partial}{\partial
      \theta}
\right ]
Y_{lm}
\\
&+&
F^{1-\ord} (l) F^{\ord-1}(l-1)
\frac{1}{2l +1}
\left [  l \cos \theta - \sin \theta \frac{\partial}{\partial
      \theta}
\right ]
Y_{lm}
\end{eqnarray}
and
\begin{eqnarray}
\label{KA2}
\fl
F^{1-\ord}(l)
\widetilde{F}^{\ord-1}
&\Big[& 
\Big (S^{+ (1)} \cos \theta \Big )
\Big (S^{-(1)} Y_{lm}
\Big )
\Big ]
=
\\ \nonumber
&-& F^{1-\ord} (l) F^{\ord-1}(l+1)
\frac{l}{2l +1}
\left [  (l+1) \cos \theta + \sin \theta \frac{\partial}{\partial
      \theta}
\right ]
Y_{lm}
\\ \nonumber
&+& F^{1-\ord} (l) F^{\ord-1}(l-1)
\frac{l+1}{2l +1}
\left [  (l+1) \cos \theta + \sin \theta \frac{\partial}{\partial
      \theta}
\right ]
Y_{lm}
\\ \nonumber
&+&
\iu  \frac {\partial}{\partial \phi}  Y_{l m}.
\end{eqnarray}
%
Finally, we obtain:
\begin{equation}
\label{boppszap}
\fl
\psF_{S_3}^{(\ord)} \star Y_{lm}
=
\Big [
\cos \theta
\coef_1^{(\ord)} (l)
+
\sin \theta
\frac {\partial}{\partial \theta}
\coef_2^{(\ord)} (l)
-
\frac {\iu}{2}
\frac {\partial}{\partial \phi}
\Big ]
Y_{lm}
\end{equation}
with $\coef_i^{(\ord)},\ i=1,2$ defined in (\ref{fs}).




\end{document}